\documentclass[twocolumn]{aastex63}

\usepackage[english]{babel}
\usepackage[T1]{fontenc}
\usepackage{textgreek}
\usepackage{amsmath}
\usepackage{graphicx}
\usepackage{commath}
\usepackage{multirow}
\usepackage{color}
\usepackage{aecompl}
\usepackage{tikz}
\usepackage{import}
\usepackage{booktabs}
\usepackage{mathtools}

\DeclareRobustCommand{\ion}[2]{%
\relax\ifmmode
\ifx\testbx\f@series
{\mathbf{#1\,\mathsc{#2}}}\else
{\mathrm{#1\,\mathsc{#2}}}\fi
\else\textup{#1\,{\mdseries\textsc{#2}}}%
\fi}

\newcommand{\I}{\protect\scriptsize I \normalsize $\!\!$}
\newcommand{\HI}{\mbox{\rm H\,\I}\: }

\definecolor{mossgreen}{rgb}{0.68, 0.87, 0.68}
\definecolor{strawberry}{rgb}{0.99, 0.42, 0.52}
\definecolor{lapislazuli}{rgb}{0.15, 0.38, 0.61}
\definecolor{pinegreen}{rgb}{0.0, 0.47, 0.44}
\usepackage[colorinlistoftodos,prependcaption,textsize=scriptsize]{todonotes}

\newcommand{\fix}[1]{\textcolor{black}{#1}}

\shorttitle{OHM Identification Methods \& New Detections} 
\shortauthors{Roberts et al.}

\submitjournal{\apj}
\accepted{2025 April 8}
\published{2025 June 10}

\begin{document}

\title{Testing OH Megamaser Identification Methods in \ion{H}{I} Surveys: Updated Source-Flagging Algorithms and New Detections in ALFALFA}

\author[0000-0003-0046-9848]{Hayley Roberts}
\affil{Center for Astrophysics and Space Astronomy, 
Department of Astrophysical and Planetary Science, 
University of Colorado, 
389 UCB, Boulder, CO 80309-0389}
\affil{Minnesota Institute for Astrophysics,
University of Minnesota, 
116 Church Street SE, 
Minneapolis, MN 55455} 
\affil{School of Physics and Astronomy, 
University of Minnesota, 
116 Church Street SE, 
Minneapolis, MN 55455} 

\author[0000-0003-2511-2060]{Jeremy Darling}
\affil{Center for Astrophysics and Space Astronomy,
Department of Astrophysical and Planetary Science,
University of Colorado, 
389 UCB, Boulder, CO 80309-0389}

\author[0000-0001-9662-9089]{Kelley M. Hess}
\affil{Department of Space, Earth and Environment, 
Chalmers University of Technology, 
Onsala Space Observatory, SE-43992 
Onsala, Sweden}
\affiliation{ASTRON, Netherlands Institute for Radio Astronomy, Oude Hoogeveensedijk 4, 7991 PD Dwingeloo, The Netherlands}

\author[0000-0002-7892-396X]{Andrew J. Baker}
\affil{Department of Physics and Astronomy, 
Rutgers, The State University of New Jersey, 
136 Frelinghuysen Road, Piscataway, NJ 08854-8019}
\affil{Department of Physics and Astronomy,
University of the Western Cape,
Robert Sobukwe Road, Bellville 7535, South Africa}

\author[0000-0002-9798-5111]{Elizabeth A. K. Adams}
\affiliation{ASTRON, Netherlands Institute for Radio Astronomy, Oude Hoogeveensedijk 4, 7991 PD Dwingeloo, The Netherlands}
\affiliation{Kapteyn Astronomical Institute, University of Groningen, Landleven 12, 9747 AD, Groningen, The Netherlands}

\author[0000-0002-9214-8613]{Helga D\'enes}
\affiliation{School of Physical Sciences and Nanotechnology, Yachay Tech University, Hacienda San Jos\'e S/N, 100119, Urcuqu\'i, Ecuador }

\correspondingauthor{Hayley Roberts}
\email{hayleyroberts.astro@gmail.com}

\begin{abstract}
OH megamasers (OHMs) are extragalactic masers found primarily in gas-rich galaxy major mergers. To date, only $\sim$120 OHMs have been cataloged since their discovery in 1982, and efforts to identify distinct characteristics of OHM host galaxies have remained inconclusive. As radio astronomy advances with next-generation telescopes and extensive 21 cm \ion{H}{I} surveys, precursors to the Square Kilometre Array (SKA) are expected to detect the 18 cm OH masing line with significantly increased frequency, potentially expanding the known OHM population tenfold. These detections, however, risk confusion with lower-redshift \ion{H}{I} emitters unless accompanied by independent spectroscopic redshifts. Building on methods proposed by \cite{Roberts2021} for distinguishing these interloping OHMs via near- to mid-IR photometry and emission line frequencies, we apply these techniques to data from the Arecibo Legacy Fast ALFA [Arecibo L-band Feed Array]  (ALFALFA) survey and a preliminary APERture Tile In Focus (Apertif) \ion{H}{I} emission line catalog from the Westerbork Synthesis Radio Telescope. Our study, utilizing the Apache Point Observatory 3.5m telescope to obtain optical spectroscopic redshifts of 142 candidates (107 from ALFALFA and 35 from Apertif), confirms five new OHM host galaxies and reidentifies two previously catalogued OHMs misclassified as \ion{H}{I} emitters in ALFALFA. These findings support the predictions from \cite{Roberts2021} and underscore the evolving landscape of radio astronomy in the context of next-generation telescopes.
\end{abstract}


\section{Introduction}

OH megamasers (OHMs) are rare luminous 18 cm masers found in (ultra-)luminous infrared galaxies ([U]LIRGs), produced in gas-rich major mergers, and are associated with extreme star formation rates \citep{Darling2007}. Despite extensive searches, only about 120 OHM hosts have been found in the 40 years since their initial discovery (e.g., \citealt{Baan1982,Darling2002c,Willett2012}; Roberts \& Darling in prep.). Most of these searches have targeted IR luminous galaxies; however $\sim80\%$ of (U)LIRGs show no masing activity \citep{Willett2011a}, which results in low OHM detection rates.  Current efforts to isolate the physical conditions responsible for producing OHMs are similarly frustrated by the small number of known OHMs. 

\cite{Briggs1998} predicted that the 18 cm OH line could ``spoof'' the 21~cm line of neutral hydrogen (\ion{H}{I}) in untargeted extragalactic surveys, leading to contamination of \ion{H}{I} catalogs. This can happen  when there is no independent redshift of a galaxy, allowing an OHM host to appear to be at a redshift of $z_\mathrm{HI}$ when the actual redshift is $z_\mathrm{OH} = \left(\nu_\mathrm{OH,0}\left(1+z_\mathrm{HI}\right)/\nu_\mathrm{HI,0}\right) - 1$, where $\nu_\mathrm{OH,0}$ is the rest frequency of OH (1667.35903 MHz) and $\nu_\mathrm{HI,0}$ is the rest frequency of \ion{H}{I} (1420.40575 MHz).  \cite{Suess2016} first demonstrated this possibility by finding five OHMs masquerading as \ion{H}{I} sources in the 40\% data release of the Arecibo Legacy Fast Arecibo L-band Feed Array (ALFALFA) survey \citep{Haynes2018}. \cite{Roberts2021} showed that future \ion{H}{I} surveys with the Square Kilometre Array (SKA) and its precursors will detect thousands of new OHMs. At high redshifts ($z\gtrapprox1.5$), OHM detections will likely outnumber \ion{H}{I} detections in flux-limited emission line surveys. Projections over the next decade indicate that the number of known OHMs could increase by an order of magnitude \citep{Roberts2024}. Recently, record-shattering studies reported the first detections of OHM hosts at a redshift of $z \geq 0.5$ \citep{Glowacki2022} and in turn at $z \geq 0.7$ \citep{Jarvis2024}. Both of these discoveries originate from preliminary \ion{H}{I} survey data taken by MeerKAT \citep{Jonas2016}, the South African SKA precursor, further supporting the predictions of an upcoming onslaught of new detections.

Although these forthcoming detections will be helpful in determining what differentiates OHM hosts from non-masing (U)LIRGs, identification of OHMs in \ion{H}{I} surveys is expected to be a challenge. Upcoming surveys will detect thousands to hundreds of thousands of \ion{H}{I}-emitting galaxies. Although some of these galaxies will have existing spectroscopic optical redshifts that allow for effective identification of potential OHMs, many will not. To address this challenge, \cite{Roberts2021} presented new machine learning methods that can identify potential OHM hosts based on near- to mid-IR photometry in the absence of independent redshift measurements. 

In this paper, we test the \cite{Roberts2021} methods on two \ion{H}{I} survey datasets: the full ALFALFA survey \citep{Haynes2018} and a preliminary catalog using new data from the Apertif Wide-area Extragalactic Survey (AWES; \citealt{Adams2022}). This exercise allows these methods to be tested on both legacy data with a large number of sources and new data with higher sensitivity. To test the algorithms, we obtained optical spectroscopic redshifts from the 3.5\,m telescope at the Apache Point Observatory (APO). These redshifts were used to determine the rest frequency of the observed radio emission line in each galaxy. The results of these verification tests and observations led to revisions in the OHM flagging algorithms presented here, as well as the identification of five new OHMs in ALFALFA.

This paper is organized as follows. In Section \ref{sec:source_selection}, we describe how the catalog of potential OHM hosts was selected and present refinements to the original algorithms developed in \cite{Roberts2021}. Section \ref{sec:observations} details our APO observations and data reduction process. We present the results of our observations in Section \ref{sec:results}. Section \ref{sec:discussion} discusses what these results tell us about the numbers of OH detections in untargeted \ion{H}{I} surveys, including two other previously missed OHM hosts in ALFALFA. A brief summary of the main points of this paper is presented in Section \ref{sec:conclusion}. Throughout this work, when calculating OH line luminosities, we assume a flat $\Lambda$CDM cosmology with $H_0 = 70$ km s$^{-1}$ Mpc$^{-1}$, $\Omega_m = 0.3$, and $\Omega_\Lambda = 0.7$.

\section{Source Selection \& Algorithm Revisions} \label{sec:source_selection}

In order to search for previously unidentified OHMs, we applied the algorithm presented in \cite{Roberts2021} to two extragalactic \ion{H}{I} catalogs: the full ALFALFA catalog \citep{Haynes2018} and a preliminary catalog from AWES using Apertif, a phased array feed for the Westerbork Synthesis Radio Telescope (WSRT; \citealt{VanCappellen2022}). 
The ALFALFA catalog consists of 31,502 extragalactic \ion{H}{I} sources out to redshift $z_\mathrm{HI}<0.06$ (or, for unrecognized OH emitters, $0.174<z_\mathrm{OH}<0.244$). ALFALFA's nearly 7,000 deg$^2$ of sky coverage includes two large areas that cover declinations of 0 to +36 degrees.
The preliminary Apertif catalog, constructed using 4.5 months of Apertif observations, consists of $\sim$1,200 \ion{H}{I} sources up to redshift $z_\mathrm{HI}<0.066$ ($0.174<z_\mathrm{OH}<0.251$), covering declinations of >27 degrees. As this catalog was preliminary, it initially included some duplicate sources and spurious detections that were removed as they were identified, which is why only an estimate for the number of sources is provided.

Both parent catalogs were cleaned of sources that had optical spectroscopic redshift-confirmed \ion{H}{I} identifications and were then cross-matched with the AllWISE (Wide-field Infrared Survey Explorer; \citealt{Wright2010}) catalog to obtain photometry in WISE bands W1, W2, and W3 (3.4, 4.6, and 12 $\mu$m, respectively). For the ALFALFA catalog, we used a $10^{\prime\prime}$ cross-match radius from the listed optical counterpart due to the lower astrometric accuracy of WISE compared to the optical imaging. Due to Arecibo's large L-band beam size, some of these optical counterparts are found to have been incorrectly identified, and for those sources, we later investigated WISE crossmatches within a radius of $30^{\prime\prime}$ of the beam center for each detection. This value is slightly larger than the reported \ion{H}{I} centroid uncertainties of 20$^{\prime\prime}$, as this positional uncertainty may be higher for low SNR sources, as discussed in \cite{Haynes2011}. For the Apertif catalog, we cross-matched each of the detected emission lines to the AllWISE catalog with a search radius of $10^{\prime\prime}$ (as allowed by the smaller WSRT beam size). \cite{Roberts2021} present two algorithms for identifying potential OHMs using WISE photometry: one that requires data from the W1 and W2 bands (the ``W1-W2 algorithm'') and one that requires data from the W1, W2, and W3 bands (the ``W1-W2-W3 algorithm''). The first algorithm was applied to the sources in the crossmatched catalogs with SNR $> 5$ for bands W1 and W2. The second algorithm was applied to sources with SNR $> 5$ for bands W1, W2, and W3 (a subset of the former). We applied the algorithms to the appropriate sub-catalogs and, for each source, estimated the probability that the putative \ion{H}{I} emitter is a misidentified OHM. We defined our starting OHM candidate catalog as containing those sources that have $> 90\%$ likelihoods of being OHMs. 


After preparing these initial catalogs, we visually inspected the sources identified as potential OHMs for both the ALFALFA and Apertif catalogs using optical images from the Sloan Digital Sky Survey (SDSS; \citealt{Stoughton2002}) and the first data release from the Panoramic Survey Telescope and Rapid Response System (Pan-STARRS; \citealt{Chambers2016}). These images are useful for assessing morphologies in our starting catalogs; OHMs are found in galaxy major mergers, so sources with undisturbed morphologies are unlikely to host OHMs and can quickly be ruled out. The initial catalogs generated by applying the original algorithms of \cite{Roberts2021} included some good examples of potential mergers or disturbed galaxies, but the selection methods also identify low surface brightness (LSB) galaxies as potential OHMs. LSB galaxies are a common but difficult-to-identify class of galaxies that have central surface brightnesses fainter than the night sky by at least one magnitude \citep{Impey1997}. They are historically defined by their $B$-band surface brightness, with typical selection criteria of $\mu_B\geq \ \sim22-23\text{ mag arcsec}^{-2}$. In optical images, they are markedly differentiated from OHM hosts because they are faint and diffuse. However, despite having low star formation rates, they are gas-rich and much more easily detectable in \ion{H}{I} emission than in optical light \citep{Du2015}. Although the exact reason that the original methods from \cite{Roberts2021} were inadvertently selecting LSB galaxies is unclear, it is likely due to multiple factors:  LSB galaxies are abundant \citep[perhaps accounting for up to \textit{half} of all local galaxies:][]{Hodges-Kluck2020}, and they have markedly lower star formation rates than the typical \ion{H}{I} emitter. Additionally, the methods of \cite{Roberts2021} are fundamentally searching for photometric outliers from a population of bright, nearby \ion{H}{I} galaxies. Although OHM hosts and LSB galaxies are distinct populations, both appear optically fainter than typical \ion{H}{I} sources -- LSB galaxies because they are nearby but faint, and OHM hosts because they are luminous but more distant.

\begin{figure}[!ht]
\centering
\includegraphics[width=0.5\textwidth]{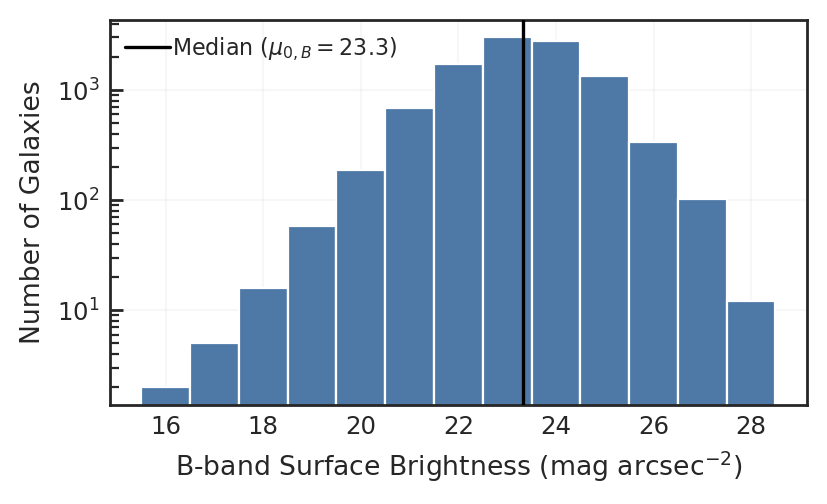}
\caption{Histogram of fit $B$-band surface brightnesses for the sample of ALFALFA galaxies that could be fit. The median value is marked with a vertical solid line.}
\label{fig:muB_hist}
\end{figure}

Although LSB galaxies are interesting in their own right, here we update the methods of \cite{Roberts2021} to properly account for and omit these sources when applying OHM selection algorithms. To update these methods, we first need to create a catalog of LSB galaxies in ALFALFA as a training sample. To do so, we calculated the $B$-band surface brightness of each source with SDSS spectroscopic redshifts and DR16 $g$- and $r$-band photometry. The $B$-band surface brightness was used so that we could apply typical LSB cutoff values to our catalog. The surface brightness $\mu_0$, in units of mag arcsec$^{-2}$, was calculated for each galaxy using the formulation presented in \cite{Du2015}:
\begin{gather}
\mu_0(m) = m + 2.5 \log_{10} (2 \pi a^2 q) - 10 \log_{10}(1+z) \\ 
\mu_0 (B) = \mu_0(g) + 0.47(\mu_0(g)-\mu_0(a)) + 0.17
\end{gather}
where $m$ refers to the apparent magnitude of the relevant band ($g$ or $r$), $a$ is the on-sky semi-major axis of the galaxy, $q$ is the axis ratio, and $z$ is the SDSS spectroscopic redshift. For the on-sky radius of the galaxy, we adopted the Petrosian radius fit to the $r$-band image in the SDSS pipeline.  We calculated the axis ratio using the de Vaucouleurs ellipticity, $\varepsilon_\mathrm{dV}$, fit to the $r$-band image where $q=1-\varepsilon_\mathrm{dV}$. Since we used the Petrosian radius, we also used the Petrosian magnitude to calculate the surface brightnesses for the $g$ and $r$ bands. For each value used from the SDSS catalog, we require a SNR of 10 to rule out any poor fits, but we still expect surface brightness values to be approximate due to the use of pipeline-fit parameters. However, this uncertainty should not be a major concern, since our goal is to roughly flag galaxies with low surface brightness (and star formation rate) and distinguish them from the more actively star-forming \ion{H}{I} and OHM hosts. A histogram of $B$-band surface brightness values is shown in Figure \ref{fig:muB_hist}. To select the sources that are considered LSB galaxies, we imposed a generous cut of $\mu_0(B)\geq23.5$ mag arcsec$^{-2}$ to ensure that we are well within the LSB galaxy parameter space. We visually inspected sources near this selection threshold to verify that most LSB galaxies were included in this sample. In some cases, sources that may not be true LSB galaxies were included as well. However, these sources are still fainter and more diffuse than typical OHM host galaxies, resulting effectively in no impact on our OHM classification algorithms.

Using our triaged sample of non-LSB \ion{H}{I} emitters, LSB \ion{H}{I} emitters, and OHM hosts, we followed a process nearly identical to that described in \citep{Roberts2021} to recreate the W1-W2 and W1-W2-W3 algorithms to identify potential OH sources in untargeted \ion{H}{I} surveys. However, this time, instead of assigning sources to one of just two classifications, we now assign them to one of three. To do so, we employ a one-vs.-all (OVA) classification scheme \citep{Rifkin2004}, which works by iterating through all classes and then defines a classifier that distinguishes sources in a given class from all sources not in that class. This approach differs from a one-vs.-one (OVO) classification scheme, which constructs classifiers for each pair of classifiers. We chose the OVA classification scheme because it is simple to interpret --- each source is assigned a likelihood of belonging to each class --- and it is computationally efficient. We also performed tests comparing the OVA and OVO classification schemes, and the OVA classification scheme yielded a higher OH recall (i.e., fewer false negatives). The obvious issue with either of these schemes is that a given source can legitimately be classified in more than one way: the LSB galaxies are, in fact, \ion{H}{I} emitters, and OHMs can also produce \ion{H}{I} emission lines (although the \ion{H}{I} emission will often not fall within the survey frequency band). 
However, the simplification of dividing our sample into three disjoint groups is helpful in creating a weighting scheme for identifying sources that are potential OHMs. Future work in this area will require a more careful examination of how to classify individual sources, but that effort will be made much more feasible with the discovery of more OHMs with more diverse host properties.




\begin{figure*}[!ht]
\centering
\includegraphics[page=3,width=0.6\textwidth]{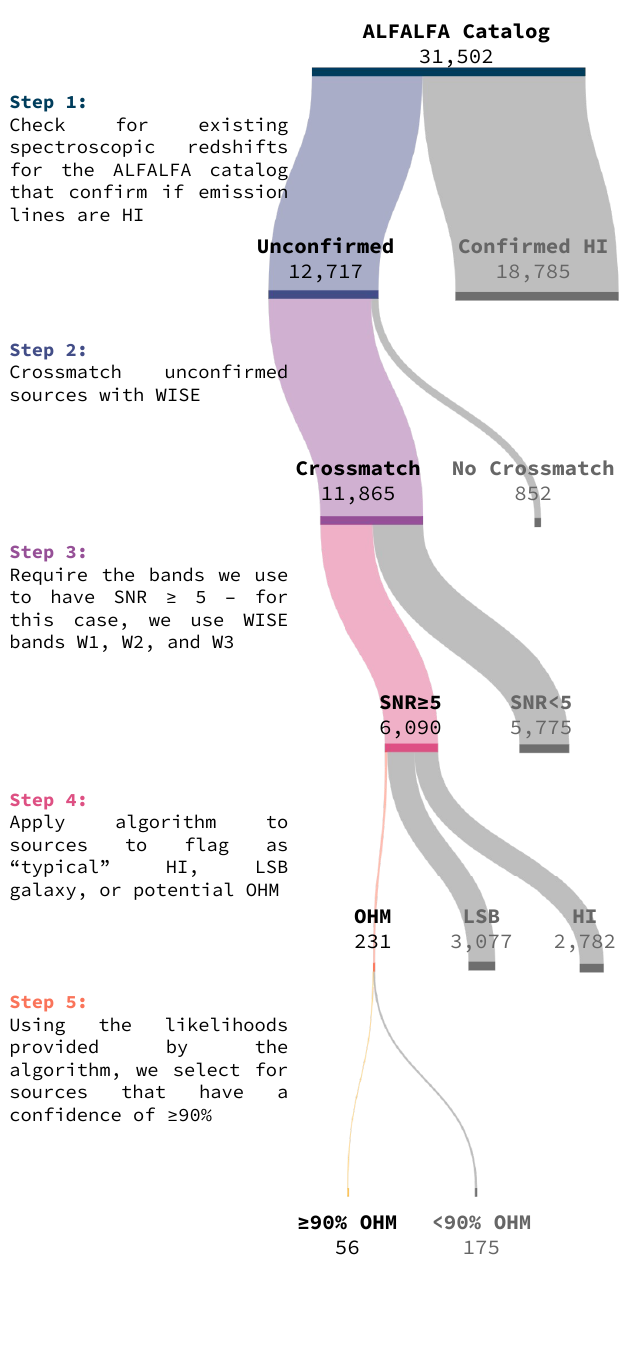}
\caption{Illustration of how sources previously identified as \ion{H}{I} emitters were selected for inspection to determine if they are misidentified OHMs. This diagram shows an example using the ALFALFA survey as input catalog and the results from the updated W1-W2-W3 algorithm that categorizes sources as ``typical'' brightness \ion{H}{I} sources (\ion{H}{I}), low surface brightness (LSB) \ion{H}{I} sources, or potential OHMs. From the potential OHM category, we primarily selected sources that have been classified with 90\% or greater likelihoods of being OHMs.}
\label{fig:W123_sankey}
\end{figure*}

Our adapted algorithms were applied to the ALFALFA-WISE and Apertif-WISE catalogs to identify potential OHMs, as described above for the original algorithms, yielding 275 high-confidence OHMs using the W1-W2 algorithm and 56 sources using the W1-W2-W3 algorithm. Figure \ref{fig:W123_sankey} shows how, starting from the original ALFALFA catalog, we drilled down to a sample of 56 for the W1-W2-W3 case. Our starting sample for optical observations was drawn from these two pools; however, since the latter observations took place over 13 months, as sources were found to be OHMs or \ion{H}{I} emitters, we fed them back into the algorithms with the correct labels, improving but potentially changing the classifications assigned to individual sources. As a result, the observing pool itself evolved, with some sources added or removed at different times as more information was taken into account. All observed sources were at some point classified as high-likelihood OHMs, but not every observed source was \textit{always} classified as a high-likelihood OHM. 
This iterative process allows the method to adapt and improve as more OHMs are discovered. 


\section{Observations \& Data Reduction} \label{sec:observations}

The identification of OHMs in \ion{H}{I} surveys relies on optical redshift measurements of host galaxies, as the detection of a single radio emission line cannot independently constrain the rest frequency (i.e., whether it is the 21cm \ion{H}{I} or the 18cm OH line). However, assuming that the emission line is \ion{H}{I} or OH, its observed frequency narrows the redshift down to two potential values. To determine whether sources are misidentified OHMs rather than \ion{H}{I} emitters, we observed our candidates to independently determine their spectroscopic redshifts using optical emission lines. We used the Dual Imaging Spectrograph (DIS) on the 3.5\,m telescope at Apache Point Observatory (APO) over 12 half-nights from December 2020 to December 2021. The same setup, presented below, was used for all 12 half-nights. Typical seeing during these observing sessions ranged from $0.7^{\prime\prime}$ to $1.4^{\prime\prime}$ except for three nights that were heavily impacted by snow and clouds, resulting in no usable data. Two other nights were affected by wind and wildfires, which limited sky areas that could be observed, but spectra were still able to be acquired. 

DIS is a medium-dispersion spectrograph with separate red and blue channels with the dichroic split occurring at 5350 \AA. Our setup consisted of R300/B400 gratings in the red and blue channels, respectively, with central wavelengths of 7500/4400 \AA\ and approximate dispersions of 2.3/1.8 \AA\ pixel$^{-1}$ spanning over 2000 pixels, or 4620/3660 \AA. The corresponding resolving power is R$\sim$3250 for the red channel and R$\sim$2400 for the blue channel.

Each night, we calibrated our data with bias frames, flat frames taken using a bright quartz lamp, and two wavelength calibrators: a Helium-Neon-Argon (HeNeAr) lamp for coarse calibration and then night sky lines for finer calibration. This calibration yielded uncertainties of $\lesssim0.73$ \AA\ and $\lesssim0.35$ \AA\ for the red and blue sides, respectively, but varied each night based on CCD fringing or observing conditions. As we only aimed to determine the redshift of each source, no flux calibrations were taken, and therefore we only report observed wavelengths of lines but not their fluxes. For each potential OHM, we took one to four five-minute exposures, depending on the brightness of each source. Some objects were revisited on multiple nights due to intermittent clouds or poor atmospheric stability. 

Data reduction was performed using the IRAF \texttt{longslit} package following typical reduction procedures. Briefly, these consisted of trimming, debiasing, and flattening science images using master biases and flats taken each night before or after observing. The science images were then wavelength calibrated using the HeNeAr images taken each night. After averaging all science images for each object, we then performed a final wavelength calibration of the averaged science images using source-free skylines. These skylines are then removed, and the final spectrum from each combined science image is extracted. We identified relevant emission lines and fit a Gaussian profile to each.

\begin{figure*}[!ht]
\centering
\includegraphics[width=0.9\textwidth]{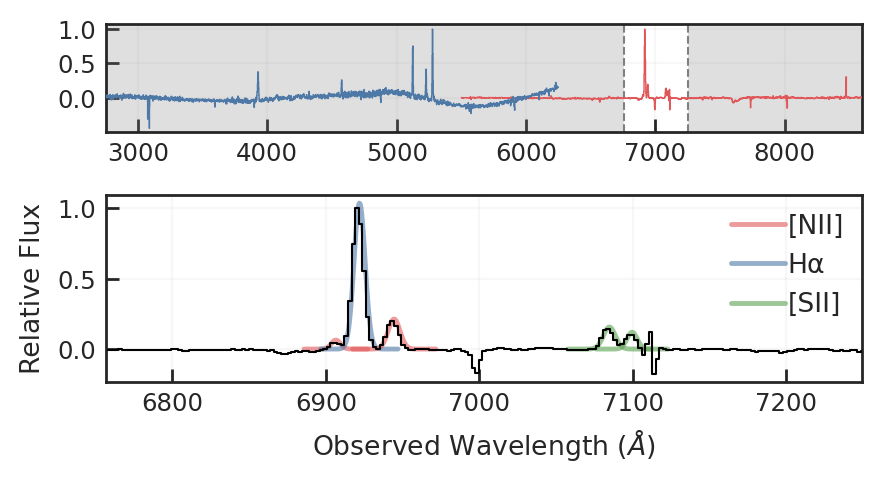}
\caption{The upper panel shows the full observed optical spectrum of AGC 720264 from both the red and blue channels. The spectral baseline oscillations in the blue spectrum result from fringing and sensitivity issues, as mentioned in Section \ref{sec:observations}; thus, all line measurements were obtained from the red channel spectra. The highlighted portion of the spectrum in the upper panel indicates the wavelength range used for line measurements, shown in the lower panel. The fits to each line are also shown as the thicker, colored overlays.}
\label{fig:example_spectrum}
\end{figure*}

Although our main objective was to identify H$\alpha$ (6563 \AA) in each spectrum, the typical spectrum had three to five emission lines, such as those of the [NII] doublet (6549 and 6583 \AA) or the [SII] doublet (6717 and 6731 \AA). Figure \ref{fig:example_spectrum} shows an example spectrum with all five of these lines and our fits to each for one of our objects. For some objects, the H$\beta$ (4861 \AA), [OIII] (4959 and 5007 \AA), or [OII] (3727 \AA) lines could be found on the blue side, but due to significant fringing and reduced sensitivity in the blue channel optics, we do not rely on these measurements for redshift calculation. The measured RMS noise for each target was calculated from source-free regions of the spectrum, and we required SNR$\geq5$ for the H$\alpha$ emission line, but typical values were SNR$\geq20$. 

Optical redshifts are calculated only on the basis of the H$\alpha$ emission because it has a significantly higher SNR and smaller measurement errors than the other lines. Other emission lines were used to confirm agreement with H$\alpha$, but due to blending and being much fainter, the [NII] and [SII] lines had much higher uncertainties and inflated the uncertainty in any redshift measurement. In an error-weighted average, they had minimal impact on the redshift. For a small number of sources, only the H$\alpha$ line is detected, and for these sources, we only present optical redshift measurement if it is in good agreement with the presumed \ion{H}{I} or OH redshift. H$\alpha$ centroiding uncertainties range from 0.001 to 0.12 \AA\ with a median value of $\sim$ 0.016 \AA . Adding the maximum calibration uncertainty (0.73 \AA\ for the red side of the spectrum) in quadrature with the maximum measurement uncertainty (0.12\AA) yields a maximum uncertainty of 0.74 \AA, corresponding to a maximum redshift uncertainty of 1.1$\times$10$^{-4}$ or about 33 km s$^{-1}$ in recession velocity. Although many of our optical redshift measurements have errors much lower than this value, we adopt it as the uncertainty for all of our measured optical redshifts.

\section{Results \& Analysis}\label{sec:results}
Each source we observed yielded one of four results:

\textit{1. The source was correctly identified as an \ion{H}{I} emitter.} This determination was made when the measured optical redshift coincided within the uncertainty with the redshift calculated assuming that the detected radio spectral line is \ion{H}{I} ($z_\mathrm{HI}$). 

\textit{2. The radio source was incorrectly identified as \ion{H}{I} and the reported spectral line is actually OH.} This determination was made when the optical redshift matched within the uncertainty the redshift calculated assuming that the detected radio spectral line is OH ($z_\mathrm{OH}$). These sources are new OHM discoveries that had masqueraded as \ion{H}{I} sources. Converting from the incorrect $z_\mathrm{HI}$ to the correct $z_\mathrm{OH}$ is done using the following equation, where the \ion{H}{I} rest frequency ($\nu_\mathrm{HI,0}$) is 1420.405752 MHz and the OH rest frequency ($\nu_\mathrm{OH,0}$) is 1667.35903 MHz:
\begin{equation}\label{zHI_to_zOH}
z_\mathrm{OH} = \frac{\nu_\mathrm{OH,0}}{\nu_\mathrm{HI,0}}(1+z_\mathrm{HI})-1.
\end{equation}

\textit{3. The source's optical redshift did not match either the \ion{H}{I} or the OH redshift.} These cases are likely the result of false positive radio spectral line detections, or the optical counterparts to the radio line have not yet been correctly identified. There is also the remote possibility that the detected line is real but neither \ion{H}{I} nor OH, such as a radio recombination line or some new line.  

\textit{4. The optical source did not produce identifiable emission lines.} Some sources, even after long integration times, did not produce identifiable optical emission lines that could be used to determine redshifts. It is possible that some of these sources are incorrectly identified optical components that had higher redshifts than could be measured by our observing setup and thus actually belong to group 3 above, but more specific determinations could not be made. We also examined the spectra for any potential absorption lines, but no decisive features were available to further classify these sources.

The results of our observation campaign are presented in Appendix \ref{sec:apo_results} with individual line measurements presented in Table \ref{tab:line_measurements}. In total, we observed 142 optical objects associated with 120 radio emission line sources. As the beam size of radio telescopes is often much larger than the optical size of the host galaxy, we sometimes observed multiple optical counterparts per radio line emitter. 
Of these optical sources, 35 were associated with radio emission lines from Apertif and 107 were associated with lines from ALFALFA.

\subsection{\ion{H}{I} Confirmations \& Ambiguous Sources}
Our observations confirmed 78 \ion{H}{I} sources -- 25 from Apertif and 53 from ALFALFA.  These sources are presented in Table \ref{tab:observing_hi} along with the corresponding optical positions associated with the matching redshifts. For sources from the ALFALFA catalog, the AGC number is provided. Optical positions were determined using SDSS DR9 images, as well as telescope-pointing images while observing. 

Table \ref{tab:observing_ambiguous} presents the optical sources that were observed where no redshift determinations could be made, either because they lacked optical emission lines to determine redshifts or because their optical redshifts matched neither the expected OH nor \ion{H}{I} redshifts. Since multiple optical sources could correspond with the radio emission line due to large beam sizes, these 60 optical sources correspond to 46 radio sources. Broken down by survey, 10 optical sources were observed corresponding to 8 radio emission lines from Apertif, and 50 optical sources corresponding to 38 radio emission lines from ALFALFA. In total, 17 optical sources were found to have redshifts that did not correspond to the expected \ion{H}{I} or OH redshifts. For another 43 sources, no redshifts could be measured from their spectra. 

\begin{deluxetable*}{lcDccc}
\tablecaption{OH line properties of new OHM confirmations\label{tab:new_ohms}}
\tablehead{
	\colhead{AGC} & \colhead{Optical Position} & \multicolumn2c{$z_\mathrm{opt}$} & \colhead{OH Line Flux} & \colhead{Line Width} & \colhead{$\log L_\mathrm{OH}$}\\
	\colhead{ } & \colhead{(J2000)} & \multicolumn2c{} & \colhead{(Jy km s$^{-1}$)} & \colhead{(km s$^{-1}$)} & \colhead{(L$_\odot$)}
}
\decimals
\startdata
102299 & 003924.7+260414.4 & 0.20310 & 0.67 & \phn69 & 2.90 \\
116345 & 011604.0+110136.6 & 0.20509 & 1.60 & 242 & 3.29 \\
193884 & 093238.3+161157.0 & 0.19095\tablenotemark{a}  & 2.30 & 454 & 3.88 \\
249507\tablenotemark{b} & 140340.3+295456.0 & 0.17862 & 1.94 & 254 & 3.26 \\
322050 & 221306.3+011627.0 & 0.18435 & 2.89 & 437 & 3.46 
\enddata
\tablenotetext{a}{This redshift was provided by SDSS.}
\tablenotetext{b}{This source was first proposed to be an OHM in \cite{Haynes2018} but had not yet been confirmed.}
\tablecomments{Flux and line width values are from \cite{Haynes2018} and are used to calculate OH luminosity. The line width is measured as $W_\mathrm{50}$.}
\vspace{-0.5cm}
\end{deluxetable*}


\subsection{OH Detections} \label{sec:OH_detections}

\begin{figure*}[!th]
    \centering
    \includegraphics[width=0.49\textwidth]{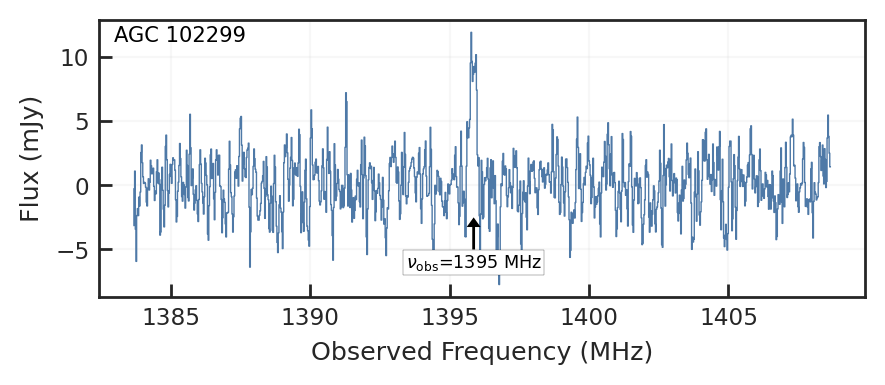}
    \includegraphics[width=0.49\textwidth]{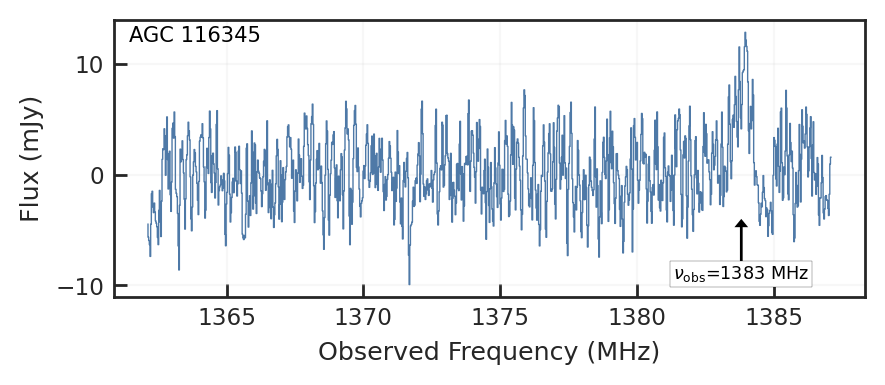}
    \includegraphics[width=0.49\textwidth]{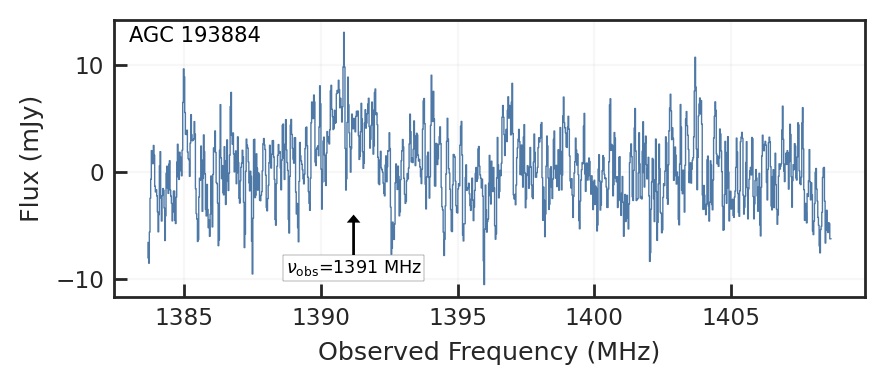}
    \includegraphics[width=0.49\textwidth]{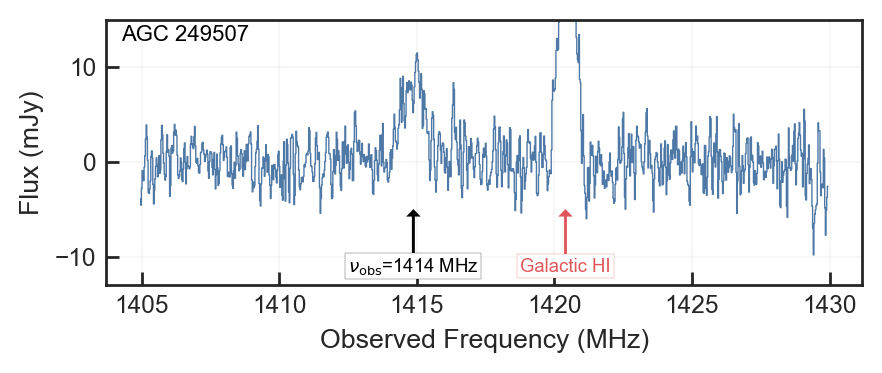}
    \includegraphics[width=0.49\textwidth]{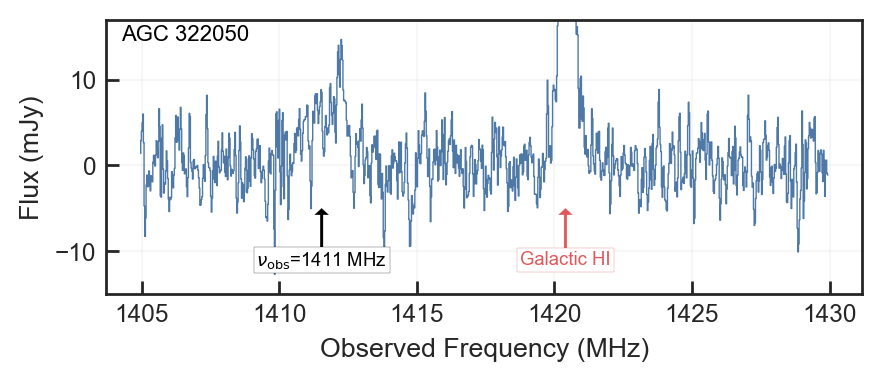}
    
    \includegraphics[width=0.49\textwidth]{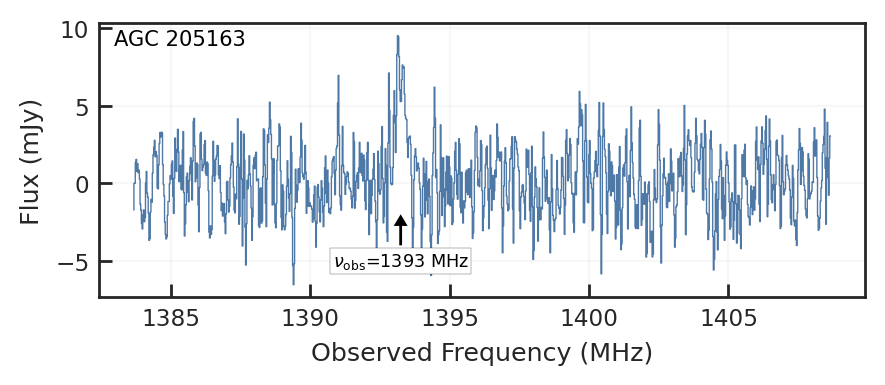}
    \includegraphics[width=0.49\textwidth]{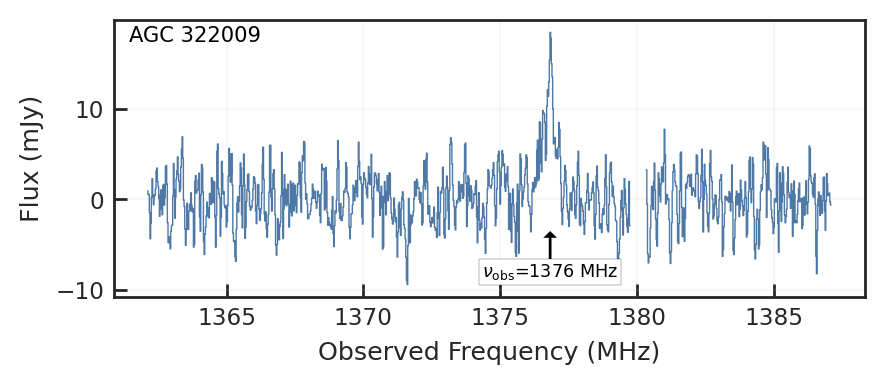}
    \caption{Spectra from ALFALFA for  seven OHMs, the top five panels showing sources identified as OHMs through spectroscopic redshifts in this work and the bottom two panels showing the misidentified sources discussed in Section \ref{sec:missed_ohms}. 
    \fix{AGC 249507 is included in the table of potential OH emission lines in \cite{Haynes2018}, for which no spectra were published. The spectrum presented here was provided by M. P. Haynes (private communication, 2024)}. Each spectrum indicates the central frequency of the OH line, as measured by ALFALFA.}
    \label{fig:oh_spectra}
\end{figure*}

Our observations yielded four OHM detections, all found in the ALFALFA catalog. We originally planned on observing AGC 193884; however, an optical redshift had already been obtained by SDSS that confirms that it hosts an OHM. We present the confirmation of five new OHMs and their properties in Table \ref{tab:new_ohms} and spectra for these OHMs, when available, are shown in Figure \ref{fig:oh_spectra}. As all five are from the ALFALFA catalog, the integrated flux and line width ($W_\mathrm{50}$) in an observed velocity frame from \cite{Haynes2018} are first converted from \ion{H}{I} to OH values with the following equations:
\begin{align}
    W_\mathrm{50,OH} &= \frac{\nu_\mathrm{OH,0}}{\nu_\mathrm{HI,0}}\times W_\mathrm{50,HI}, \label{eq:W50OH} \\
    S_\mathrm{OH} &= \frac{S_\mathrm{HI}}{W_\mathrm{50,HI}}\times W_\mathrm{50,OH}, \label{eq:SOH}
\end{align}
where $\nu_\mathrm{HI,0}$ and $\nu_\mathrm{OH,0}$ are the rest frequencies for each line and $S$ is the line flux in Jy km s$^{-1}$. We then use these values, as well as Equation 37 from \cite{Meyer2017} for converting line fluxes measured in the observed optical frame, to calculate the OH luminosity:
\begin{equation}\label{eq:LOH}
L_\mathrm{OH} = 4\pi D_L^2 \  \frac{\nu_\mathrm{OH,0}}{c\, (1+z)^2} \ S_\mathrm{OH},
\end{equation}
where $\nu_\mathrm{OH,0}$ is the rest frequency of the OH line
, $D_L$ is the luminosity distance and $S_\mathrm{OH}$ is the line flux in Jy km s$^{-1}$.

\cite{Haynes2018} present nine confirmed OHM detections -- some confirmed using SDSS redshifts and some previously identified by \cite{Suess2016}, \cite{Morganti2006}, and \cite{Darling2002c} --  and ten OHM candidates that require confirmation. We observed three of the OHM candidates that they presented and confirm that one of them is an OHM (AGC 249507 -- see Table \ref{tab:new_ohms}). For the other two OHM candidates we observed, AGC 749309 and 219835, the optical sources of the emission lines could not be identified, and they remain ambiguous. Together with the six confirmed OHMs from \cite{Suess2016} and the additional five OHMs discussed in \cite{Haynes2018}, this brings the total number of OHM detections in ALFALFA to 16.

\begin{figure*}[!th]
\centering
\includegraphics[width=\textwidth]{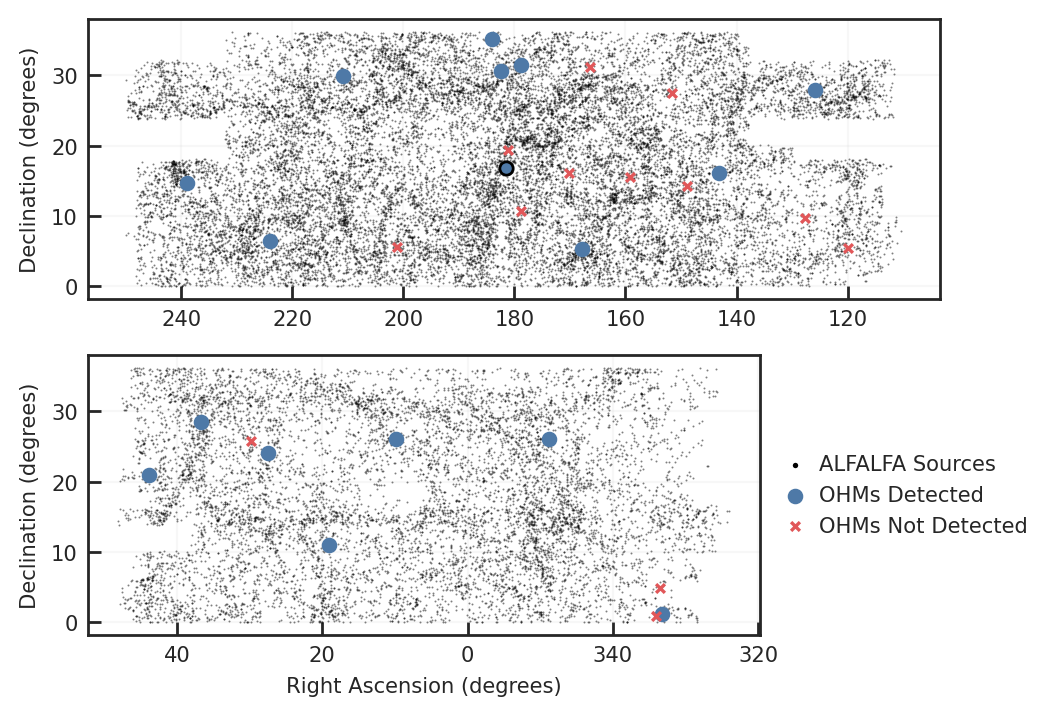}
\caption{Sky distribution of ALFALFA sources (black dots) including OHMs that were detected in ALFALFA (blue circles) and those that fall within the survey parameters but have no reported detection from ALFALFA (red x's). \fix{The marker with the black outline is IRAS 12032+1707, a known OHM host whose OH emission was not reported in \cite{Haynes2018}, but was confirmed to be detected in the ALFALFA survey by M. P. Haynes (private communication, 2024).}}
\label{fig:ALFALFA_footprint}
\end{figure*}

\section{Discussion} \label{sec:discussion}

The discovery of additional OHMs in the ALFALFA catalog is exciting but potentially not unexpected. \cite{Suess2016} previously searched for OHM candidates in the 40\% survey and confirmed six of those sources as OHMs. While not all sources determined to be OHMs in this work were available in the 40\% catalog, AGC 102299 and AGC 193884 were available in both the early iterations and final catalog. Furthermore, in the data reduction process for ALFALFA, each source was inspected to omit false positive \ion{H}{I} detections. The OHMs presented in this work evaded identification in these processes, with the exception of AGC 249507, whose proposed identification as an OHM in \cite{Haynes2018} was further supported by compact radio emission reported in \cite{Wang2024}. Although it is not clear why these sources were not previously flagged as potential OHM hosts, we can compare the number of known OHMs in ALFALFA with the predicted value to assess the likelihood of finding these sources and more.

\subsection{Comparison with the Predicted Number of OHM Detections in ALFALFA}

The total number of OHMs predicted to be detected in ALFALFA has varied: \cite{Giovanelli2005} predicted that ALFALFA would detect a few dozen OHMs, while \cite{Suess2016} extrapolated the number of OHMs detected in the ALFALFA 40\% data release to predict that roughly 15 total new OHMs would be detected throughout the ALFALFA survey. Now, with complete ALFALFA survey parameters, we can integrate over the updated Markov chain Monte Carlo (MCMC) fit to the OH luminosity function (OHLF) from \cite{Roberts2021} to calculate the number of OHMs predicted to be detected by ALFALFA. We used the ALFALFA survey parameters from \cite{Haynes2018}: a frequency range of 1435-1350 MHz\footnote{The full frequency coverage of ALFALFA is 1435-1335 MHz. However, as noted in \cite{Haynes2018}, ALFALFA suffers from significant radio frequency interference below 1350 MHz, and volume-complete studies should be restricted to volumes corresponding to radio frequencies $>$ 1350 MHz.} ($z_\mathrm{OH}$=0.162--0.235), 5$\sigma$ survey sensitivity of 0.72 Jy km s$^{-1}$ for line width $W_{50}=200$ km s$^{-1}$, and sky coverage of almost 7,000 deg$^2$ to obtain a predicted total number of OHM detections of $N_\mathrm{OH}=35.8^{+6.6}_{-6.5}$. 

\subsubsection{Misidentified OHM Hosts} \label{sec:missed_ohms}

The discrepancy between this predicted value and the 16 known OHMs discussed in Section \ref{sec:OH_detections} warrants a closer look at the ALFALFA catalog. Comparing a comprehensive OHM catalog with the ALFALFA source catalog, we find two OHMs that are detected in ALFALFA but are presumed to be \ion{H}{I} sources. We note that these two sources were flagged by our OHM finding methods; this result is expected, as both sources were known OHMs used to train our algorithms originally. IRAS 10339+1548 (denoted AGC 205163 in the ALFALFA catalog) is a known OHM host first identified by \cite{Darling2001}. The emission line detected by ALFALFA occurs at $\nu_\mathrm{obs}=1393$ MHz and, assuming the line is the previously detected OHM emission, yields a redshift of $z=0.1969$, consistent with the measured spectroscopic redshift of the host galaxy \citep{Spoon2022}. The other OHM misidentified in the ALFALFA catalog is IRAS 22135+0043 (AGC 322009), which was first identified by \cite{Willett2012}. The ALFALFA catalog gives an observed emission line at $\nu_\mathrm{obs}=1376$ MHz for this source. Assuming that it is OH instead of \ion{H}{I}, this yields a redshift of $z=0.2110$, consistent with other spectroscopic redshift measurements \citep{Glikman2018}. This brings the total number of confirmed OHMs in the ALFALFA catalog to 18.

\subsubsection{Missing OHM Detections}

\begin{figure*}[!ht]
\centering
\includegraphics[width=\textwidth]{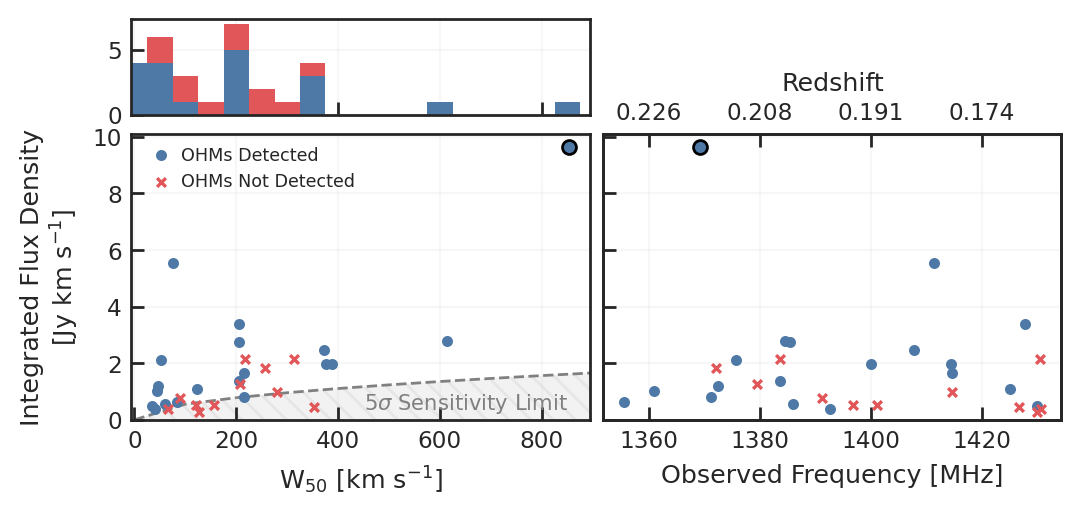}
\caption{The lower left panel shows integrated flux density vs. line width (W$_{50}$) of OHMs with reported detections in ALFALFA (blue circles) and those without reported detections (red x's). The $5\sigma$ survey sensitivity (grey dashed curve) is 0.846 Jy km s$^{-1}$ for $W_{50}=235$ km s$^{-1}$. This is rescaled for OH from the reported \ion{H}{I} sensitivity of 0.72 Jy km s$^{-1}$ for $W_{50}=200$ km s$^{-1}$ \citep{Haynes2018}. The upper left panel displays a stacked histogram of line width values. The panel on the right shows integrated flux density vs. observed frequency of the emission line, shown on the bottom, or source redshift, shown on the top. The marker with the black outline is IRAS 12032+1707, a known OHM host whose OH emission was not reported in \cite{Haynes2018}, but was confirmed to be detected in the ALFALFA survey by M. P. Haynes (private communication, 2024).}
\label{fig:ALFALFA_SdV_vs_dV}
\end{figure*}

In addition to these misidentified sources, we can inspect the catalog of known OHMs to select for any sources that fall within the footprint and frequency range of ALFALFA. Specifically, we select OHMs that fall within one of the two regions covered by ALFALFA: $7\mathrm{h}30\mathrm{m}<\mathrm{R.A.}<16\mathrm{h}30\mathrm{m}$ and $22\mathrm{h}<\mathrm{R.A.}<3\mathrm{h}$, with both regions covering a declination range of $0^\circ<\mathrm{dec.}<36^\circ$. We also require that OHMs have observed emission lines between $1335-1435$ MHz. This exercise yields an \textit{additional} 12 OHMs that fall within the footprint of the ALFALFA survey, shown as red x's in Figure \ref{fig:ALFALFA_footprint}. Determining whether these OHMs should have been detected by ALFALFA is more difficult. Figure \ref{fig:ALFALFA_SdV_vs_dV} shows the integrated flux density versus line width for OHMs with ALFALFA detections and those without reported detections. ALFALFA reports a $5\sigma$ sensitivity limit of 0.72 km s$^{-1}$ for a line width of $W_{50}=200$ km s$^{-1}$. This \ion{H}{I} sensitivity limit can be rescaled for OH using Equations \ref{eq:W50OH} and \ref{eq:SOH}, yielding a sensitivity limit of 0.845 km s${-1}$ for a line width of $W_{50, OH}=235$ km s$^{-1}$. The dashed line in Figure \ref{fig:ALFALFA_SdV_vs_dV} indicates this rescaled 5$\sigma$ survey sensitivity extrapolated across the entire range of line widths. 
We utilize this reported sensitivity to illustrate the distribution of sources with reported detections in Figure \ref{fig:ALFALFA_SdV_vs_dV}, but we note that ALFALFA required a signal-to-noise ratio of 6.5 (or 5 for those with confirmed spectroscopic redshift measurements) for inclusion in their catalog \citep{Haynes2018}. This sensitivity can also be impacted by other factors, such as radio frequency interference and the standing waves it can produce. 
Although there is no correlation between the observed frequency of a line and the likelihood that a detection was made in the ALFALFA survey, as shown in the left panel of Figure \ref{fig:ALFALFA_SdV_vs_dV}, intermittent interference could still impact detectability. Furthermore, OHMs can be variable, particularly in their peak emission \citep{Darling2002a}, which could cause them to elude detection by ALFALFA.

As some OHM emission lines can be significantly broadened, the detection of a line with high integrated flux density spread over a large number of spectral channels can be challenging. Although none of our possible explanations fully explain the distribution of OHMs that have reported detections in ALFALFA and those that do not, this last consideration is illustrated by sources with narrower line widths in Figure \ref{fig:ALFALFA_SdV_vs_dV} being more likely to have reported detections in ALFALFA. It could also be that marginally detected emission lines were not recognized as having the expected optical counterparts (spiral galaxies) and were therefore rejected as spurious during ALFALFA quality control. 

These complications make it impossible to determine which OHMs \textit{should} have been detectable by ALFALFA. However, for argument's sake, if all of these additional 12 OHMs were detected by ALFALFA, the total number of OHMs detected by ALFALFA would rise to 30. Furthermore, if the nine additional OHM candidates from \cite{Haynes2018} can be confirmed, they would increase the number of OHMs in ALFALFA to 39. Both of these values, although somewhat hypothetical, are consistent with the predicted value of $N_\mathrm{OH}=35.8^{+6.6}_{-6.5}$.

This approximate analysis further emphasizes the need for a better understanding of OHMs, their hosts, and how they can be identified. Upcoming \ion{H}{I} surveys on next-generation telescopes will significantly increase the number of known OHM hosts and allow for some of the work necessary to better understand this population. However, these surveys will also have issues similar to those that impact the search for OHMs within the ALFALFA survey. The increase in the number of detected sources will make wading through OH false positives more time-consuming and further stress the need for optical or IR spectroscopic redshifts. As surveys reach higher redshift ranges, the available ancillary data for host galaxies dwindle significantly. Tackling these challenges will require revisiting techniques, such as those presented in this work, and adapting our algorithms or developing new ones as our understanding evolves.

\subsection{Comparison with the Predicted Number of OHM Detections in Apertif}

We cannot calculate the expected number of OHM detections in the preliminary Apertif catalog using the same method as above, as the observations are only partially complete. However, we can use the expected OH contamination rate for Apertif of $0.09^{+0.02}_{-0.01}$\% from \cite{Roberts2021} and apply it to our catalog of $\sim$1,200 sources. This gives an expected number of OHM detections for the preliminary catalog of $1.1^{+0.2}_{-0.1}$ 
detections. Given the detection of the previously known OHM IRAS 10597+5926 in Apertif \citep{Willett2012,Hess2021}, the lack of additional OHMs in the preliminary catalog is consistent with the expected OHM contamination rate, validating the estimate presented in that paper.

\section{Conclusion}\label{sec:conclusion}

In this work, we tested the OHM flagging methods presented in \cite{Roberts2021} on both existing \ion{H}{I} survey data from ALFALFA and new survey data from AWES. Through these tests, we identified a tendency to flag LSB galaxies as potential OHMs. Our methods were updated to identify potential OHM hosts, accounting for the large number of LSB galaxies in \ion{H}{I} survey data. For these potential OHM hosts flagged in \ion{H}{I} surveys, we obtain longslit optical spectra in order to determine the rest frequency of the radio emission line detected in each source. In total, we obtained 142 optical spectra, confirming \ion{H}{I} emission in 78 galaxies. For 60 of these spectra, the results were ambiguous, where either a redshift could not be determined or the measured redshift did not match the redshift inferred from the \ion{H}{I} or OH line. Lastly, we were able to identify five new OHMs previously thought to be \ion{H}{I} sources in ALFALFA data. These sources verify the ability of our algorithms to successfully identify potential OHMs that were interloping in \ion{H}{I} survey data. We also identify two known OHMs that were misidentified in the ALFALFA data as \ion{H}{I} sources.

We also examine the number of expected OHM detections for both the ALFALFA and the preliminary Apertif catalogs. Although this exercise is somewhat ambiguous for a partial Apertif catalog, it shows that there are likely still some unidentified OHMs within the ALFALFA catalog, either from the list of potential OHMs from \cite{Haynes2018} or otherwise eluding any suspicion. While this scenario may seem unlikely, it is supported by numerous studies that have closely scrutinized the ALFALFA survey data for OHM hosts, only to be followed up by subsequent studies identifying previously missed OHMs (\citealt{Haynes2011,Haynes2018,Suess2016}; this work; and certainly future additional studies).

The full scope of OH contamination in \ion{H}{I} surveys is, in some ways, unknown. However, the methods presented here will continue to become more accurate as new OHMs are identified and our methods adapted as new information is obtained. For a fraction of the new OHMs that will be detected in next-generation \ion{H}{I} surveys, existing optical redshifts will be crucial for fast OHM identification and then used to further strengthen these algorithms for sources without existing redshifts. The future of OHM science will be driven by these detections in \ion{H}{I} surveys.

\acknowledgments
We are grateful to Martha P. Haynes for her generosity in searching through past ALFALFA data and sending detailed information and spectra that were essential for this paper. We thank the wonderful staff at Apache Point Observatory for observing assistance and support, in addition to excellent company. Finally, we thank the anonymous reviewer for their thorough and helpful comments which improved the presentation of this paper. This work has been supported by the National Science Foundation through the grant AST-1814648. HR acknowledges partial support from NASA Award \#80NSSC20M0057. AJB acknowledges support from the National Science Foundation through grants AST-1814421 and AST-2308161 and from the Radcliffe Institute for Advanced Study at Harvard University. This research has made use of the NASA/IPAC Extragalactic Database (NED), which is funded by the National Aeronautics and Space Administration and operated by the California Institute of Technology.

\software{Astropy \citep{AstropyCollaboration2022}, Astroquery \citep{Ginsburg2019b}, IRAF \citep{Tody1986}, Matplotlib \citep{Hunter2007}, Numpy \citep{Harris2020}, Scipy \citep{Virtanen2020}, Scikit-Learn \citep{Pedregosa2011}}

\facilities{ARC, Arecibo, WSRT}

\appendix

\restartappendixnumbering

\section{Results of Optical Redshift Observing Campaign} \label{sec:apo_results}
We present the full results from the optical redshift observing campaigns. All tables are published in their entirety in machine-readable formats, and portions of each are shown here as a demonstration. Table \ref{tab:observing_hi} presents the sources that were confirmed to be \ion{H}{I} detections. Table \ref{tab:observing_ambiguous} presents the optical sources where either no redshift measurement was made or the measured redshift was not consistent with either an OH or an \ion{H}{I} identification. 
Lastly, Table \ref{tab:line_measurements} presents all emission lines measured for optical sources where redshift determinations were made. The measurements presented here are the observed central wavelengths for each line and the uncertainties on those fits. The uncertainties are strictly statistical and do not include systematic or calibration uncertainties, as discussed in Section \ref{sec:observations}.

\begin{deluxetable*}{cccccch}
\tablecaption{\HI confirmations from APO observing campaign\label{tab:observing_hi}}
\tablehead{
    \colhead{Radio Position} & \colhead{AGC} & \colhead{$z_\mathrm{HI}$} & \colhead{$z_\mathrm{OH}$} & \colhead{Optical Position} & \colhead{$z_\mathrm{opt}$} \\
    \colhead{(J2000)} & \colhead{}  & \colhead{}    & \colhead{}    & \colhead{(J2000)} & \colhead{}        
}
\tablecolumns{6}
\startdata
001137.3+253334.0 & 102682 & 0.03494 & 0.21487 & 001137.3+253334.0 & 0.03475 & \HI \\
001231.0+183339.0 & 104713 & 0.02445 & 0.20256 & 001231.0+183339.0 & 0.02462 & \HI \\
001306.1+090959.0 & 103613 & 0.03562 & 0.21567 & 001306.1+090959.0 & 0.03561 & \HI \\
001723.2+155705.0 & 101185 & 0.02566 & 0.20398 & 001723.2+155705.0 & 0.02574 & \HI \\
003054.1+102432.0 & 105317 & 0.03656 & 0.21678 & 003054.1+102432.0 & 0.03666 & \HI \\
\multicolumn{6}{c}{\textit{Full table available online}} \\
\enddata
\tablecomments{Table \ref{tab:observing_hi} is published in its entirety in machine-readable format. A portion is shown here for guidance regarding its form and content.}
\end{deluxetable*}

\begin{deluxetable*}{ccccccl}
\tablecaption{Ambiguous sources from APO observing campaign\label{tab:observing_ambiguous}}
\tablehead{
    \colhead{Radio Position} & \colhead{AGC} & \colhead{$z_\mathrm{HI}$} & \colhead{$z_\mathrm{OH}$} & \colhead{Optical Position} & \colhead{$z_\mathrm{opt}$} & \colhead{Category} \\
    \colhead{(J2000)} & \colhead{}  & \colhead{}    & \colhead{}    & \colhead{(J2000)} & \colhead{} & \colhead{}       
}
\tablecolumns{7}
\startdata
001137.3+253334.0 & 102682 & 0.03494 & 0.21487 & 001133.2+253343.5 & \nodata & No Lines \\
004550.2+011916.0 & 103454 & 0.03633 & 0.21650 & 004550.2+011916.0 & 0.10667 & Neither\tablenotemark{a} \\
005040.4+030544.0 & 103359 & 0.04153 & 0.22261 & 005040.4+030544.0 & \nodata & No Lines \\
011604.4+110138.0 & 116345 & 0.02644 & 0.20490 & 011604.4+110138.0 & 0.02599 & Neither\tablenotemark{b} \\
230032.3+304212.0 & \nodata & 0.00268 & 0.17701 & 230032.3+304212.0 & \nodata & No Lines \\
\multicolumn{6}{c}{\textit{Full table available online}} \\
\enddata
\tablenotetext{a}{While optical redshifts are provided for these sources, we note that the redshift is calculated from only one emission line, resulting in a large uncertainty.}
\tablenotetext{b}{While these redshifts were close to the expected redshifts, other sources were found closer to the radio source and with better matched redshifts. These sources may possibly be contributing to the radio signal but are not expected to be the primary emitters.}
\tablecomments{Ambiguous sources were categorized into two groups: those whose optical spectra showed no identifiable emission lines that could be used to determine redshifts (No Lines) and those for which redshifts could be determined but did not match the expected OH or \ion{H}{I} redshifts (Neither). Table \ref{tab:observing_ambiguous} is published in its entirety in machine-readable format. A portion is shown here for guidance regarding its form and content.}
\end{deluxetable*}

\begin{deluxetable*}{cccccc}
\centerwidetable
\tablecaption{Optical emission line center measurements for candidate OHMs\label{tab:line_measurements}}
\tablehead{
	\colhead{Optical Position} & \colhead{NII} & \colhead{H$\alpha$ } & \colhead{NII} & \colhead{SII} & \colhead{SII} \\
	\colhead{(J2000)} & \colhead{(6549 \AA)} & \colhead{(6563 \AA)} & \colhead{(6583 \AA)} & \colhead{(6717 \AA)} & \colhead{(6731 \AA) }
}
\tablecolumns{6}
\startdata
001137.3+253334.0 & \nodata & 6790.836$\pm$0.027 & \nodata & 6949.352$\pm$0.068 & \nodata \\
001231.0+183339.0 & \nodata & 6724.385$\pm$0.003 & 6745.107$\pm$0.018 & 6882.014$\pm$0.013 & 6895.573$\pm$0.029 \\
001306.1+090959.0 & \nodata & 6796.524$\pm$0.018 & \nodata & 6954.956$\pm$0.054 & \nodata \\
001723.2+155705.0 & \nodata & 6731.734$\pm$0.040 & \nodata & \nodata & \nodata \\
002936.3+270147.0 & \nodata & 7567.133$\pm$0.011 & 7589.448$\pm$0.027 & \nodata & \nodata \\
\multicolumn{6}{c}{\textit{Full table available online}} \\
\enddata
\tablecomments{Table \ref{tab:line_measurements} is published in its entirety in machine-readable format. A portion is shown here for guidance regarding its form and content.}
\end{deluxetable*}

\bibliography{library}

\end{document}